\documentclass[10pt,conference,letterpaper]{IEEEtran}
\IEEEoverridecommandlockouts
\usepackage{cite}
\usepackage{tikz}

\usepackage{amsmath,amssymb,amsfonts}
\usepackage{algorithmic}
\usepackage{graphicx}
\usepackage{cleveref}
\usepackage{textcomp}
\usepackage{xcolor}
\usepackage[ruled]{algorithm2e} 
\usepackage[nolist]{acronym}
\usepackage{soul}
\usepackage{marginnote}
\usepackage{url}
\usepackage[inline]{enumitem}
\usepackage{tabularx}
\usepackage{amsthm}

\newcommand{\myurl}[1]{\texttt{#1}}

\newcommand{\myssec}[2]{%
  \subsection{#1}%
  \label{sec:#2}%
}
\newcommand{\rsec}[1]{%
  Sec.~\ref{sec:#1}%
}

\newcommand{\removed}[1]{%
  {\color{red}#1%
  }%
}
\renewcommand{\removed}[1]{}

\newcommand{\myfigeps}[3][width=0.5\textwidth]{%
\begin{figure}[tbp]%
\centering%
\includegraphics[#1]{figures/#2}%
\caption{#3}%
\label{fig:#2}%
\end{figure}%
}

\newcommand{\myfigfulleps}[3][width=\textwidth]{%
\begin{figure*}[tb]%
\centering%
\includegraphics[#1]{figures/#2}%
\caption{#3}%
\label{fig:#2}%
\end{figure*}%
}

\newcommand{\rfig}[1]{Fig.~\ref{fig:#1}}

\newenvironment{myinlinelist}%
{%
\begin{enumerate*}[label=(\roman*)]%
}%
{%
\end{enumerate*}%
}

\newenvironment{myitemlist}%
{%
\begin{itemize}[parsep=0em,leftmargin=*,label={--}]%
}%
{%
\end{itemize}%
}

\newenvironment{myenumlist}%
{%
\begin{enumerate}[parsep=0em,leftmargin=*,label=\arabic*.]%
}%
{%
\end{enumerate}%
}

\begin{document}

\title{
  In-Network Computing with FaaS at the Edge
}

\author{
\IEEEauthorblockN{Claudio Cicconetti}
\IEEEauthorblockA{\textit{IIT, CNR} --
Pisa, Italy \\
c.cicconetti@iit.cnr.it}
\and
\IEEEauthorblockN{Marco Conti}
\IEEEauthorblockA{\textit{IIT, CNR} --
Pisa, Italy \\
m.conti@iit.cnr.it}
\and
\IEEEauthorblockN{Andrea Passarella}
\IEEEauthorblockA{\textit{IIT, CNR} --
Pisa, Italy \\
a.passarella@iit.cnr.it}
}

\author{Claudio~Cicconetti,
        Marco~Conti,
        and Andrea~Passarella%
\IEEEcompsocitemizethanks{\IEEEcompsocthanksitem All the authors are with the Institute of Informatics and Telematics (IIT) of the National Research Council (CNR), Pisa, Italy.}%
}

\IEEEtitleabstractindextext{%
\begin{abstract}
  Offloading computation from user devices to nodes with processing
capabilities at the edge of the network is a major trend
in today's network/service architectures.
At the same time, serverless computing has gained a huge traction among
the cloud computing technologies and has, thus, promoted the adoption
of Function-as-a-Service (FaaS).
The latter has some characteristics that make it generally suitable to
edge applications, except for its cumbersome support of stateful applications.
This work is set to provide a broad view on the options available for
supporting stateful FaaS, which are distilled into four reference
execution models that differ on where the state resides.
While further investigation is needed to advance our understanding of the
opportunities offered by in-network computing through stateful FaaS,
initial insights are provided by means of a qualitative analysis of the
four alternatives and their quantitative comparison in a simulator.
\end{abstract}

\begin{IEEEkeywords}
  Edge computing, Serverless, Function-as-a-Service, Distributed computing, In-network intelligence
\end{IEEEkeywords}%
}

\maketitle

\begin{tikzpicture}[remember picture,overlay]
\node[anchor=south,yshift=10pt] at (current page.south) {\fbox{\parbox{\dimexpr\textwidth-\fboxsep-\fboxrule\relax}{
  \footnotesize{
     \copyright 2022 IEEE.  Personal use of this material is permitted.  Permission from IEEE must be obtained for all other uses, in any current or future media, including reprinting/republishing this material for advertising or promotional purposes, creating new collective works, for resale or redistribution to servers or lists, or reuse of any copyrighted component of this work in other works.
  }
}}};
\end{tikzpicture}%

\IEEEdisplaynontitleabstractindextext

\IEEEpeerreviewmaketitle

\bibliographystyle{unsrt}


%
  \section{Introduction}%
  \label{sec:introduction}%
  Today we are witnessing the transition of service provisioning from
cloud-centric to edge-centric~\cite{Campbell2019}:
%
%
centralizing storage and processing in data centers is now showing
its limitations for a wide class of scenarios where a significant
amount of traffic is generated/consumed at the edge of the network,
or which include applications that cannot tolerate the latencies
associated to reaching the cloud~\cite{Ramachandran2021}.
%
%
All these applications require offloading (part of) their computation tasks
to external elements with processing capabilities, e.g., because their local
resources are CPU- or memory-limited or to save power since they have constrained
energy availability~\cite{Liu2020}.
Furthermore, they do not always need long-term storage of their transactions.
Thus, in-network processing is a much better option than centralizing computation
in a cloud platform, since this keeps processing close to where data are
actually used, which can be also an additional benefit in terms of
privacy~\cite{Sapio2017}.

Serverless computing is thriving among cloud providers as the next step of
evolution from \ac{IaaS} and \ac{PaaS}.
With serverless, the service providers
are offered an extremely light abstraction of the physical computation resources:
they simply need to upload a container image, while
orchestration, monitoring, and scaling are handled by the
platform~\cite{Castro:2019:RSC:3372896.3368454}.
%
%
%
All serverless platforms support a programming model called
\textit{\ac{FaaS}}, where the basic unit of computation is the
\textit{function}: developers can focus on the implementation of
elementary units with well-defined input vs. output, while providers
compose such functions into chains of invocations to offer
complex services to the end users~\cite{Baldini2017}.
In principle, due to its programming simplicity, \ac{FaaS}
is a good candidate for in-network processing at the
edge of the network, and some studies show promising
results in this direction~\cite{Cicconetti2020}.
%
%
However, many practical applications are made of \textit{stateful} tasks.
In~\cite{Bhasi2021} the authors discuss the issues they had to
face in the migration of some micro-service-based applications to serverless,
which required defining \textit{ad hoc} solutions for state management.
Examples of applications include a real-time collaborative \LaTeX\
editor, where the state is the document being edited, and an
e-commerce application, where the state is the list of items in the
basket.
In \ac{IoT} applications, which are even more relevant at the edge,
commonly the tasks have a strong dependency from one another as
they compose \ac{ML} pipelines, where the state are the models or
windows of observations~\cite{Rausch2021}.
%
To manage the state, cloud serverless platforms include backend services for
state synchronization and data exchanges between function
execution~\cite{Mvondo2021}: however, the use of these services
incurs additional costs, increases vendor lock-in, and violates the
underlying pure functional paradigm.
Indeed, efficient state management has been identified as a critical open issue
in the position paper~\cite{Khandelwal2021}.
At the edge, the above issues are exacerbated by the decentralized nature of
computation elements over geographical distances, the limited capabilities of
edge nodes (compared to high-end servers in a data center), and the possible
presence of multiple competing platform providers in the same
area~\cite{Aslanpour2021}.

In this paper we first very briefly survey serverless works in the
literature that are particularly relevant to aspects related to
edge computing systems (\rsec{soa}).
We then focus on the execution of stateful functions at the edge, for which
we propose four models in \rsec{contribution}, which are compared via
simulation in \rsec{eval}.
This work sets the scene for further research activities on the
emerging topic of in-network computing at the edge through
serverless/FaaS, as discussed in \rsec{conclusions}.
  \section{Related work}%
  \label{sec:soa}%
%
%
Serverless computing only recently began expanding towards edge deployments
(see~\cite{Aslanpour2021}).
%
In addition to a more efficient execution of functions on edge
devices, which have more limited capabilities than their cloud
counterparts
, deploying at the edge requires modifications to the system architecture.
For instance, in~\cite{Ascigil2021} the authors investigate the
possibility to decentralize control to local edge networks, as
opposed to manage resources in a fully centralized manner.
Also the problem of allocating applications, consisting of multiple
inter-related functions, to edge nodes has been studied in some
works, for instance, in~\cite{Wang2020b} the authors propose a
mathematical formulation of the problem taking into account different
cost categories (activation, placement, proximity, sharing), for
which they put forward offline and online heuristic algorithms.

Another key research challenge is how to support data-intensive applications:
in~\cite{Rausch2021} the authors focus on selecting the best edge nodes for the
execution of functions that match the application requirements and data
locations.
In this work, instead, we explore the concept of stateful \ac{FaaS}
in an open manner, and envision the generic mechanisms to support
stateful serverless applications at the edge.
We have taken inspiration from~\cite{9193994}, where we have
proposed to dispatch stateless functions using lightweight brokers installed
at each edge node, and~\cite{Baresi2019}, which defines an architecture
where stateless micro-services can be deployed anywhere in the device-to-cloud
continuum, while stateful components operate in the device or the cloud only.





%


%

%
  \section{Execution models of stateful FaaS}%
  \label{sec:contribution}%
  In this section, we introduce
some terminology and fundamental concepts of cloud serverless.
Afterwards, we explain in \rsec{cont:serverless-edge} how the
fundamental design principles of cloud serverless are violated at the edge, and
introduce four execution models of stateful \ac{FaaS}
(\rsec{cont:execution}), with a focus on execution of \acp{DAG}
applications, see \rsec{cont:dags}.

\myssec{Serverless in the cloud}{cont:serverless-cloud}

\myfigeps{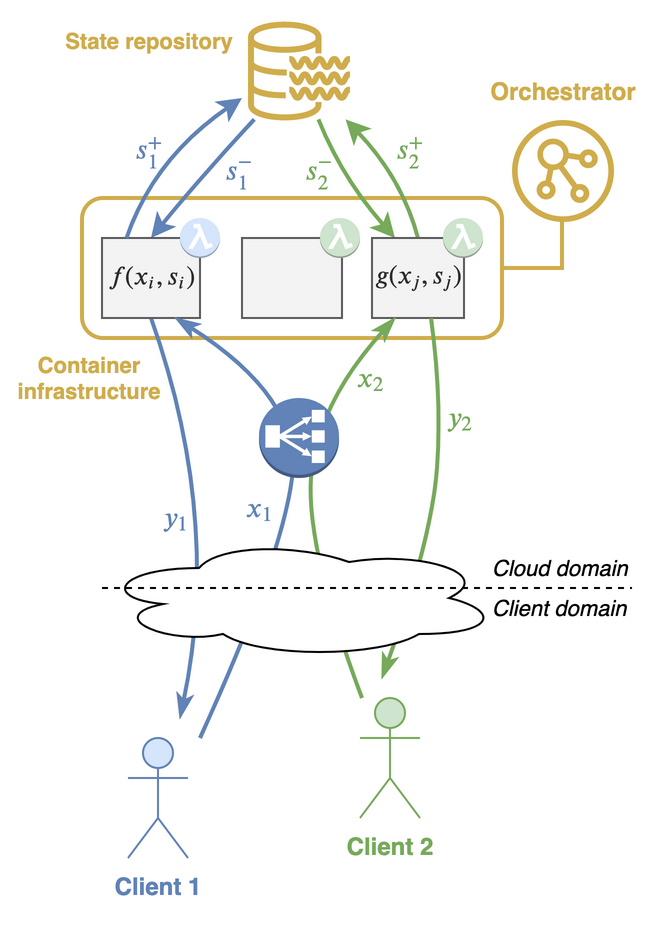}{FaaS execution model in serverless cloud.}

The typical execution model of \ac{FaaS} in cloud serverless is illustrated
in \rfig{serverless}.
%
%
The clients invoke the execution of functions (e.g., $f(\cdot)$ and $g(\cdot)$
in the figure) by providing their input $x_i$ and expecting a return value $y_i$.
All the invocations are directed towards a logically centralized entry point,
which performs load balancing towards one of the \textit{workers}, i.e.,
instances of containers of matching type in a virtualized infrastructure,
managed by an orchestrator.
%

A function may require accessing the state of an application:
this is done through an external service providing persistence, such as a
database; in our figure this is represented
by the execution of $f(x_i,s_i)$ and $g(x_j,s_j)$, where $s_i$ and $s_j$ are the states
bound to the application in execution at clients $i$ and $j$, respectively.
Commercial cloud service providers encourage using services in the
same ecosystem to disguise stateful operations as stateless, since
this increases billing and vendor lock-in.
This model highlights the following implicit assumptions of
serverless/\ac{FaaS}:

\begin{myenumlist}
    \item\label{itm:assumption1}
    \textit{Access to the state is cheap:} in commercial systems
    the state repository in \rfig{serverless} is located
    in the same data center as the serverless platform, and
    the state usually consists of a small amount
    of data that can be retrieved/updated with a single query.

    \item\label{itm:assumption2}
    \textit{Location of workers is irrelevant:} since all the
    workers run in containers managed by the same orchestration
    platform, the physical location of a worker in general
    can be assumed to have a negligible impact on the performance:
    in~\cite{Mohan2019} the authors
    have found that cold-start effects
    are dominated by the creation and initialization of
    containers' namespaces, rather than other location-dependent effects.

    \item\label{itm:assumption3}
    \textit{Location of clients is irrelevant: } as shown in
    \rfig{serverless}, there is a full separation between the clients
    (in the Internet) and the serverless platform services (in the data
    center), their only point of contact being the load balancer, which
    is logically centralized and located within the cloud domain.

\end{myenumlist}

\myssec{Serverless at the edge}{cont:serverless-edge}

\myfigfulleps{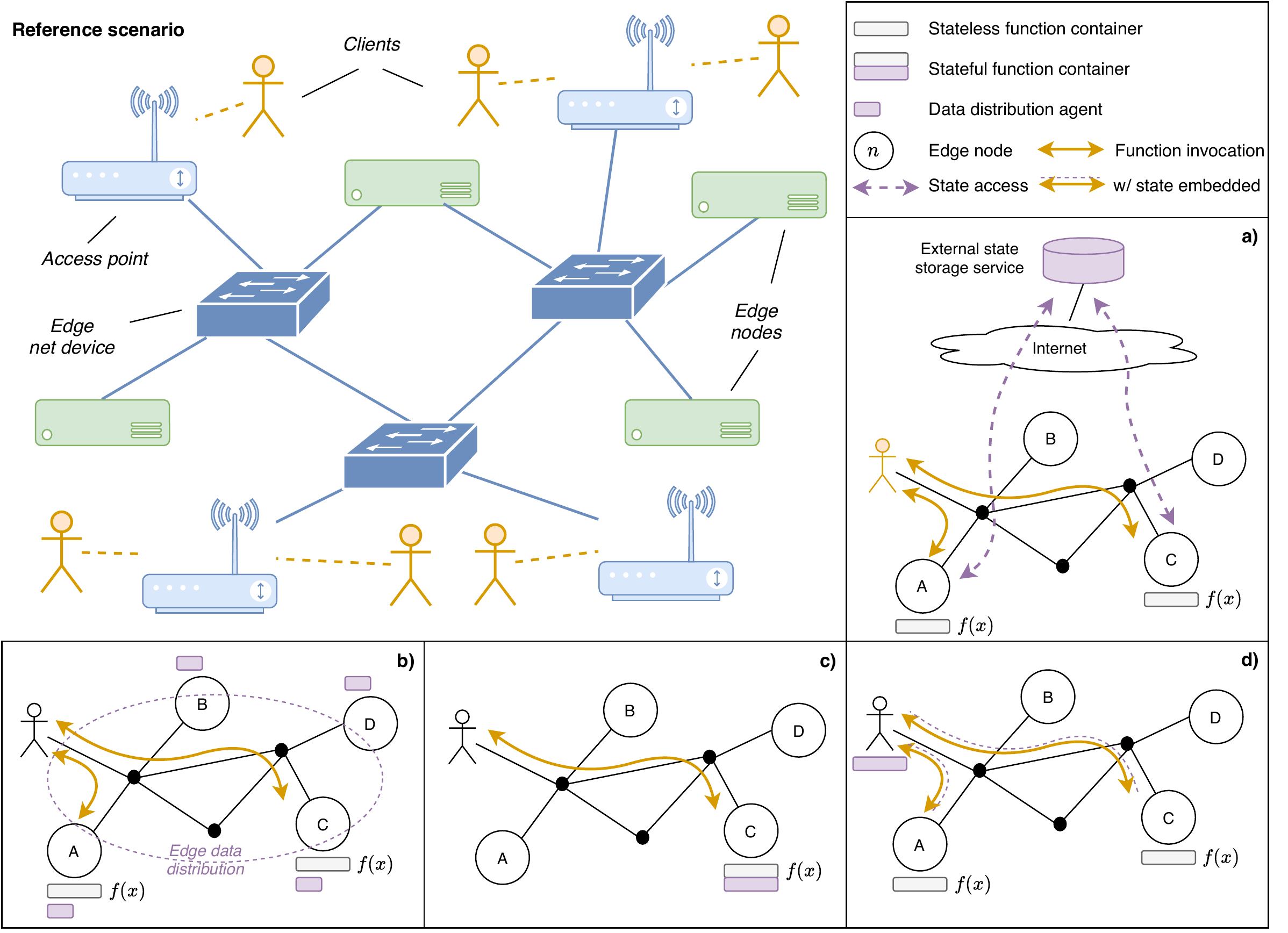}{Four stateful FaaS execution models in serverless edge:
a) external;
b) in-edge;
c) in-function;
d) in-client.}

%
A reference scenario of edge computing is illustrated in the top left 
part of \rfig{models}: \textit{edge nodes} have compute capabilities
that can be used to provide services
to \textit{clients}, which reach them
via access networks created by \textit{access points} and internal
connectivity provided by \textit{edge network devices}.
The scenario may represent a smart city with sensing nodes or an
industrial \ac{IoT} deployment with embedded devices with constrained
resources.

Some of the edge elements may have an Internet connection,
but, in general, reaching the cloud cannot be considered to be \textit{cheap} like for
cloud serverless.
In fact, one of the driving motivations of edge computing is that
it reduces the application latency since computation happens close
to the clients: if the edge nodes have to reach the cloud to
retrieve/update the application state as part of the function
invocation procedure, then the edge advantage may simply vanish
(\textit{violation of assumption \ref{itm:assumption1}}).

Furthermore, edge networks may be very heterogeneous, in terms of both
the compute capabilities of the nodes (ranging from single-board computers
to full-fledged servers) and the communication links: therefore,
the function invocation performance depends very much on the location
of the worker, unlike what happens in a cloud deployment
(\textit{violation of assumption \ref{itm:assumption2}}).

Finally, in edge networks there is no clear
separation between clients and processing nodes:
the users are interspersed in the same network connecting the
edge nodes to one another.
As a result, the location of the clients matters a lot in determining
the overall quality of the performance they perceive, because the
cost to reach a node is not uniform across all the clients (\textit{violation
of assumption \ref{itm:assumption3}}).
%
%

In summary, \textit{none of the design assumptions of cloud serverless
(\rsec{cont:serverless-cloud}) hold at the edge}: this is particularly
problematic for stateful functions, since
they are less flexible regarding their lifecycle and incur
an additional overhead to access the state.

\myssec{Stateful FaaS execution models}{cont:execution}

In the following we illustrate four models for the execution of
stateful \ac{FaaS} in an edge network, with different characteristics.
This analysis is a step forward towards the definition
of better solutions to support (stateful) \ac{FaaS} at the edge under
different application constraints and deployments: this is
an open area of investigation, whose importance is steadily
increasing thanks to the widespread diffusion of edge computing and the
growing appeal of serverless for not only mobile but also \ac{IoT}
and data analytics applications.

The four models are shown with the help of the example in \rfig{models}.
The first model (\rfig{models}a) is a straightforward transposition of
cloud serverless to an edge network: the workers remain
in fact stateless, as they access the state on demand via external
services in the cloud.
We thus call this model \textbf{external}.
It has the advantage that a function developed for the cloud can be
deployed at the edge without any modifications.
However, there is a cost for reaching the state repository in the cloud,
which has to be paid in terms of application latency and outbound traffic.

Our second model, called \textbf{in-edge} (see \rfig{models}b),
is similar to the previous one, but with an important difference: the
functions remain stateless, but the state-associated data are kept
by the edge nodes themselves.
This removes the economic/performance burden of accessing external services,
but requires \textit{data distribution agents} to be deployed
on heterogeneous and resource-constrained nodes, possibly with limited
connectivity.
An in-edge model is proposed, for instance, by
Cloudstate (\myurl{https://cloudstate.io/}), which relies on a commercial system
(Akka) to manage the applications' state in a distributed manner.
%
%
However, to the best of our knowledge, a general-purpose definitive
solution has not yet been found.

A different approach is shown in \rfig{models}c, where the
state of a function for a given application instance is kept within
the worker itself, called \textbf{in-function}.
In \ac{FaaS}, the developer is not allowed to use local resources
to keep persistent data, whereas with this model such a feature
must be provided by the serverless platform and used by the programmer.
In this model there is a mapping between a given application context and its
worker, which outlasts a single invocation: we represent this
in the figure with a single container being present in the edge network,
towards which all functions invocations of that client must be directed.
The in-function model is expected to perform better than the others, because it does
not require any transfer of information for the function to retrieve/update
the state and it allows exploiting data locality (e.g., hot caches).
%
%
The in-function model can be realized via the integration of a
system-wide serverless framework with an underlying Kubernetes
(K8s), as in~\cite{Rausch2021}, which proposes to influence the
scheduling of K8s by adjusting its internal weights based on metadata
specified by the application developers.
%

Finally, in \rfig{models}d we illustrate
\textbf{in-client}, where the state is embedded in the function arguments
(so that it can be read by the container executing the function) and
in the return value (so that a modified state can be returned to the client).
The client remains the sole owner of the state, whose ownership is delegated
only temporarily for the execution of a single function invocation.
This way, the containers are in fact stateless, as in external and
in-edge, and consecutive invocations of the same function from the
same client may happen on different edge nodes, as shown in the example.
The traffic never leaves the edge network,
but in-client may increase the amount of internal traffic generated because
the state has to be piggybacked on every function invocation exchange,
which is instead unnecessary with the other models.
Furthermore, the serverless framework has to provide a means for the
client to embed the state in the arguments and return value.

To summarize:

\begin{myitemlist}
    \item All the models identified, except in-function,
    can be used jointly with scheduling policies to dispatch the function
    invocation requests towards any of the currently available containers
    that can serve requests of matching types: this can be exploited to
    follow short-scale variable load/network conditions as in~\cite{9193994}.

    \item In-function, instead, uses stateful containers.
    Therefore, it must rely on
    container migration to optimize the performance by
    adapting to the changing environment; migration may incur
    a non-negligible cost in terms of unavailability, negatively impacting the
    quality perceived by the application~\cite{Puliafito2020}, and hence
    cannot be done too frequently.
    
    \item In-client is the only model offering a straightforward
    opportunity to enable multi-tenant applications: since the client
    owns its data, it is possible to use multiple service providers at
    the same time, which might coexist in the same geographical area
    or even in the same virtualized infrastructure, thus contributing
    to the removal of dreaded vendor lock-ins.
\end{myitemlist}

\myssec{Execution of DAGs}{cont:dags}

\myfigeps[width=2.25in]{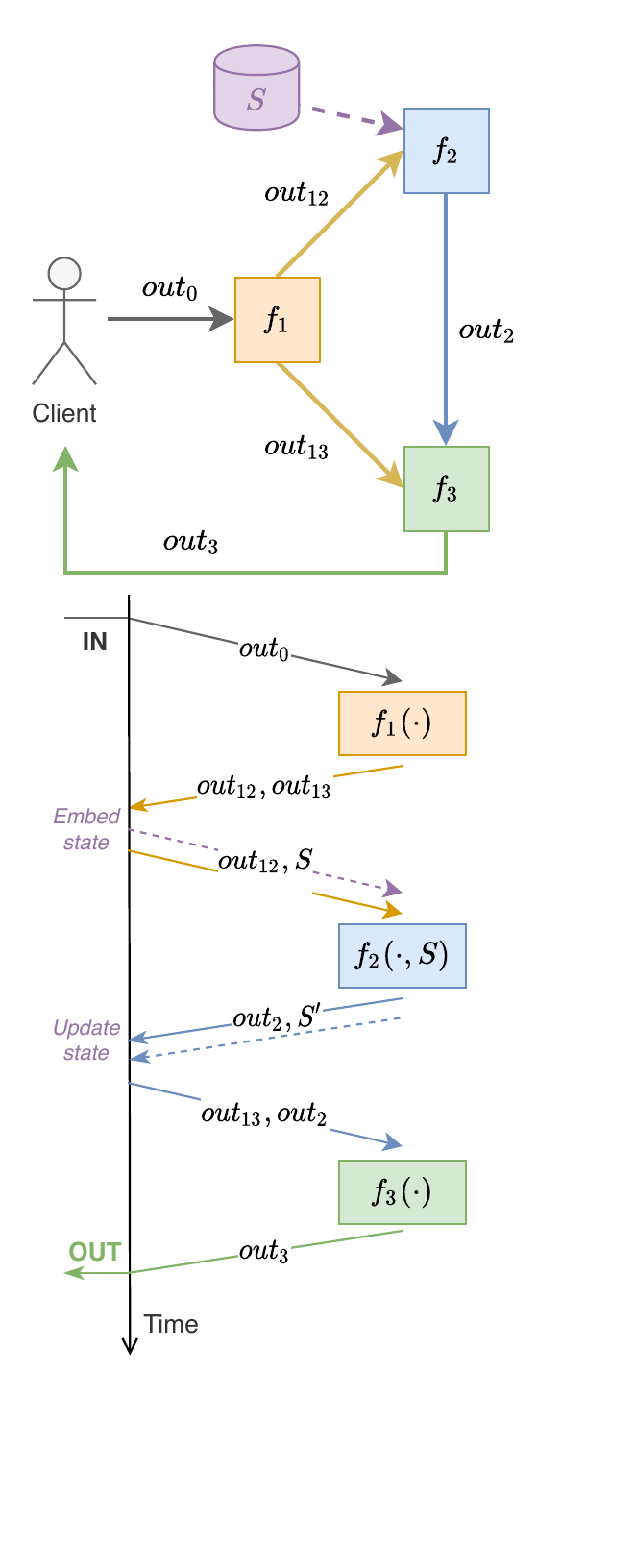}{
    Example of in-client FaaS execution of an
    application composed of three functions in a DAG, where only $f_2$
    requires access to the application's state.}

One of the most appealing features of \ac{FaaS} is that it allows
the creation of complex workflows by combining functions, so that the
input of a function is the output of one or more preceding functions
arranged in a \ac{DAG}.
An example is shown in the top part of \rfig{sequence}, where the
final output to the user is given by $f_3(\cdot)$, which takes as input
the output of $f_1(\cdot)$ and $f_2(\cdot)$, the latter also depending on
the persistent state $S$ bound to the specific application instance.
%
%

All the models in \rsec{cont:serverless-edge}, illustrated in \rfig{models},
can be adopted in a straightforward manner 
by a serverless platform that allows the composition of
functions in \acp{DAG}.
For the in-client model, we show in the bottom part of \rfig{sequence}
the sequence of function invocation exchanges to realize the example \ac{DAG}
at the top of the figure.
As can be seen, when the client invokes $f_2$ it embeds the state $S$
in the function arguments together with $out_{12}$, i.e., the output of
$f_1$ intended for $f_2$.
  \section{Performance evaluation}%
  \label{sec:eval}%
  We have carried out a high-level performance evaluation of the execution
models described in \rsec{contribution}.
Due to the absence of publicly available data from real edge
deployments, we have created our dataset for the simulation using
the technique of workload composition:

\begin{myitemlist}
    \item for the network we have used
    ``ether: Edge Topology Synthesizer'', with a
    pre-configured topology model for an \ac{IIoT} scenario
    \cite{Rausch2021}, including a mix of hardware architectures
    and communications links;

    \item for the workload we have started from \ac{DAG} traces
    generated by ``Sp\r{a}r: Cluster Trace Generator''~\cite{Tian2019},
    which we have transformed into a linear chain of stateful
    function invocations, with CPU/memory requirements and hardware
    architecture affinity.
\end{myitemlist}

Our methodology follows a Monte Carlo approach: for a given set of
configuration parameters we extract 200 random subsets of \textit{jobs}, each made
of a chain of heterogeneous function invocations with different length,
and perform an assignment of function invocations to the edge nodes
with processing capabilities.
The assignment is done using a greedy online algorithm that selects for
every function of every job (sorted randomly) the edge nodes that minimizes
the response time, according to one of two \textit{assignment policies}:
\begin{myinlinelist}
\item processing time only (\texttt{Load}), vs.
\item processing time + transfer time for input arguments, return values, and
states (\texttt{Load+net}).
\end{myinlinelist}
Such assignment assumes global knowledge of the network and task execution times.
We then measure the response time, due to processing and network transfer,
of each chain of function invocations, with the four execution models
in \rsec{cont:serverless-edge}.
For the \textit{external} model we assume that access to the
cloud has an RTT of 120~ms and a bandwidth of 50~Mb/s~\cite{Persico2017};
for the \textit{in-edge} model, for simplicity we did not simulate a
distributed storage system, but rather we assumed that the state
repository is located in the central node of the topology.
%
%
The simulator used has been released as \textit{open source} and
the instructions to fully reproduce the results are available on
GitHub (\myurl{ccicconetti/serverlessonedge}, tag \texttt{v1.1.1},
simulations \texttt{002}).
%

\myfigeps{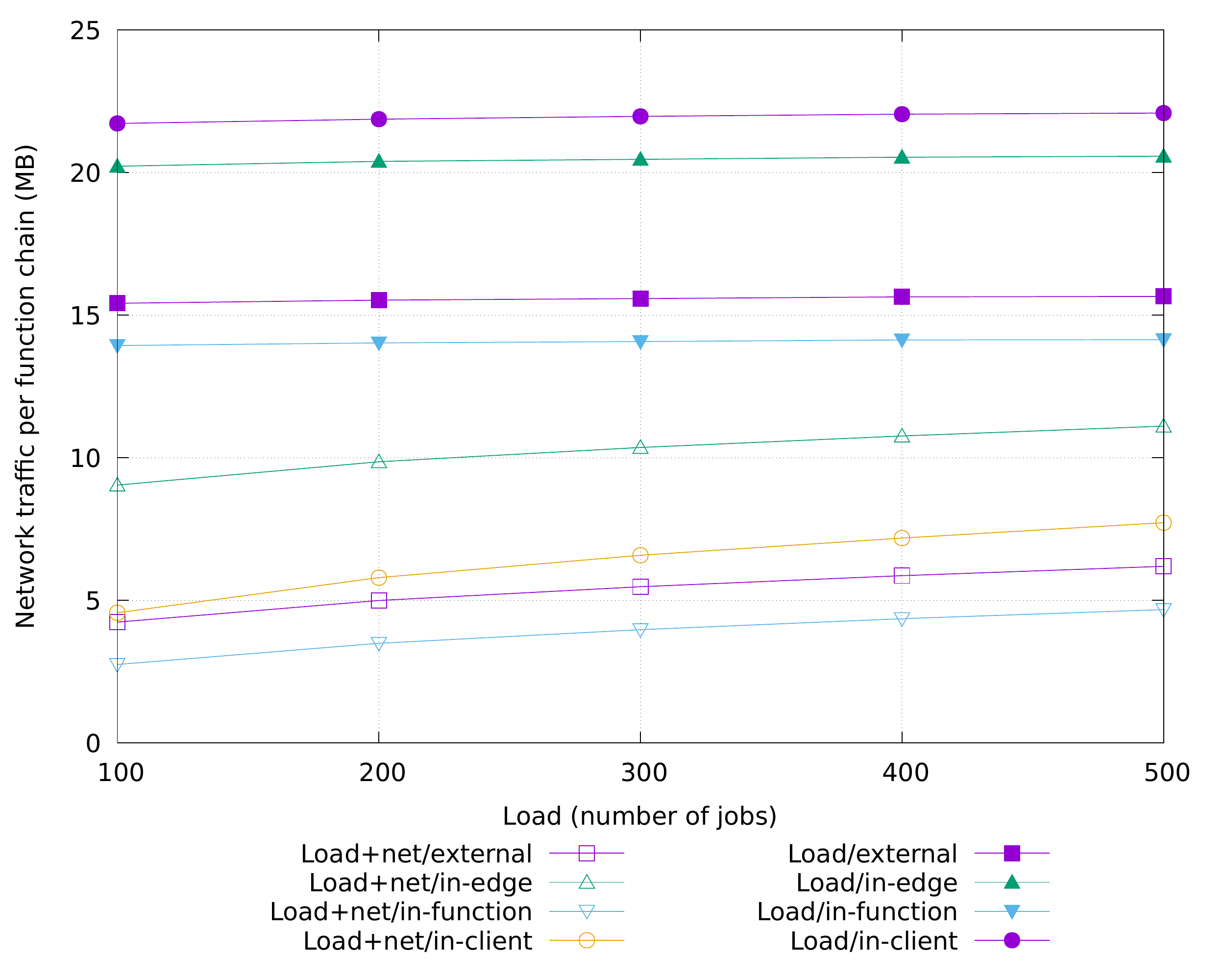}{Network traffic generated per function chain,
with both allocation strategies and all execution models, and increasing load.}

In \rfig{002-traffic-iiot} we show the traffic generated per function
chain with the four models as the load increases from 100 to 500
jobs, with the two function assignment policies.
With \texttt{Load}, \textit{in-client} execution incurs greatest
network overhead, because the functions tend to be allocated always towards
the center of the network, which has most powerful servers.
However, with \texttt{Load} the traffic is much higher than that
with \texttt{Load+net} for all execution models, which may jeopardize
the edge network resources and also negatively
impact the quality experienced by the applications.
Hence, in the following we focus on the \texttt{Load+net} allocation strategy
only.
With the latter, the traffic increases slightly with the load, and
\textit{in-edge} exhibits the worst performance, while all the other execution
models have similar performance.

\myfigeps{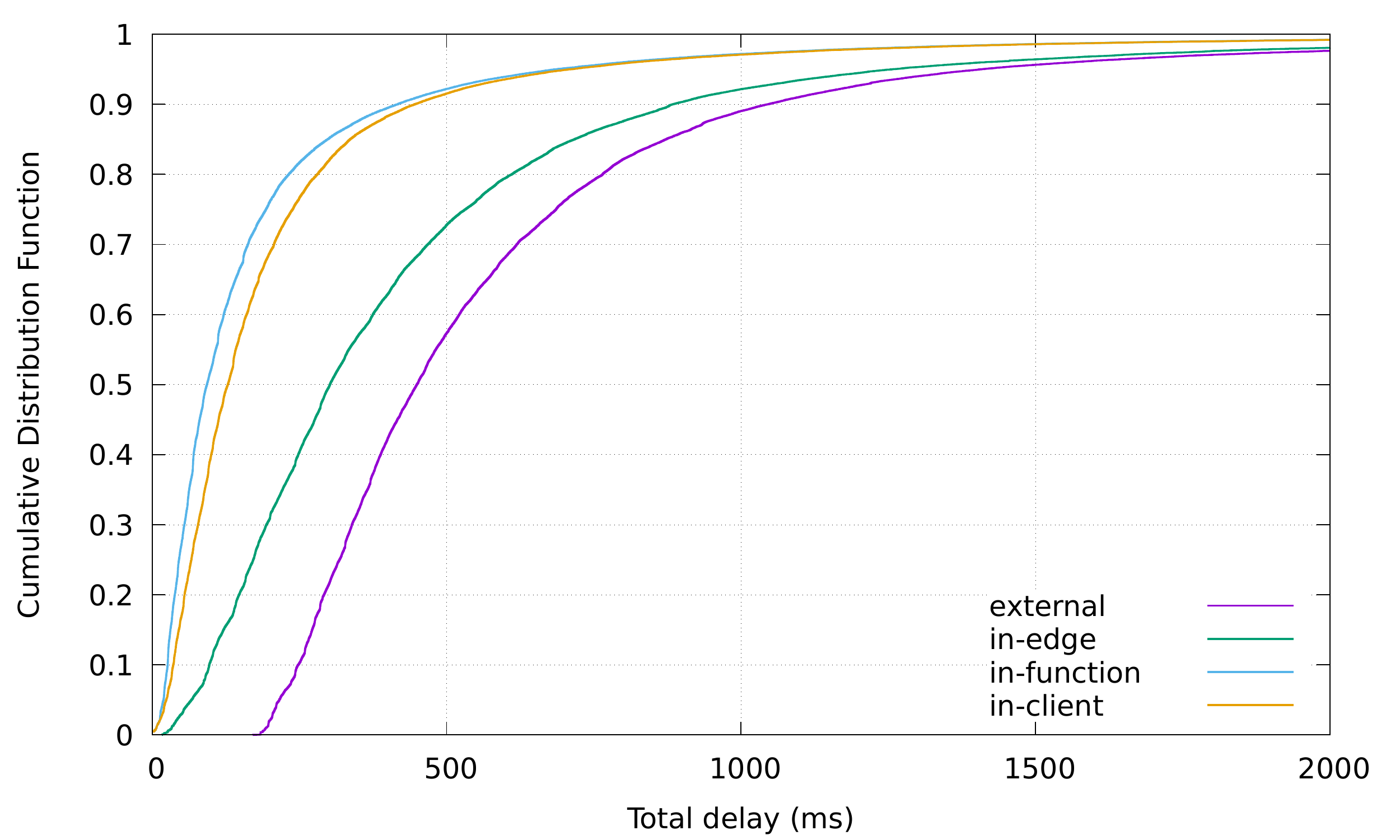}{Total latency (processing and network),
with Load+net allocation strategy, all execution models, and 500 jobs.}

For the representative case of \texttt{Load+net} and 500 jobs, we
report in \rfig{002-latency-iiot-procnet-500} the total latency of all the jobs
across all the random subsets, sorted and normalized in $[0,1]$.
As can be seen, \textit{external} and \textit{in-edge} yield a much
higher latency than the others, with \textit{in-function} performing
only slightly better than \textit{in-client}.
The results show that \textit{in-function} achieves best performance, which
however comes at the cost of adding protocol complexity and limiting
the opportunities for platform optimization, as explained in \rsec{cont:serverless-edge}.
On the other hand, even though \textit{in-client} incurs
additional overhead due to the need of piggybacking the required state
of a function on the input arguments and return value, in our
scenario the 
overall function response times are smaller than with
\textit{external} and \textit{in-edge}.

\myfigeps{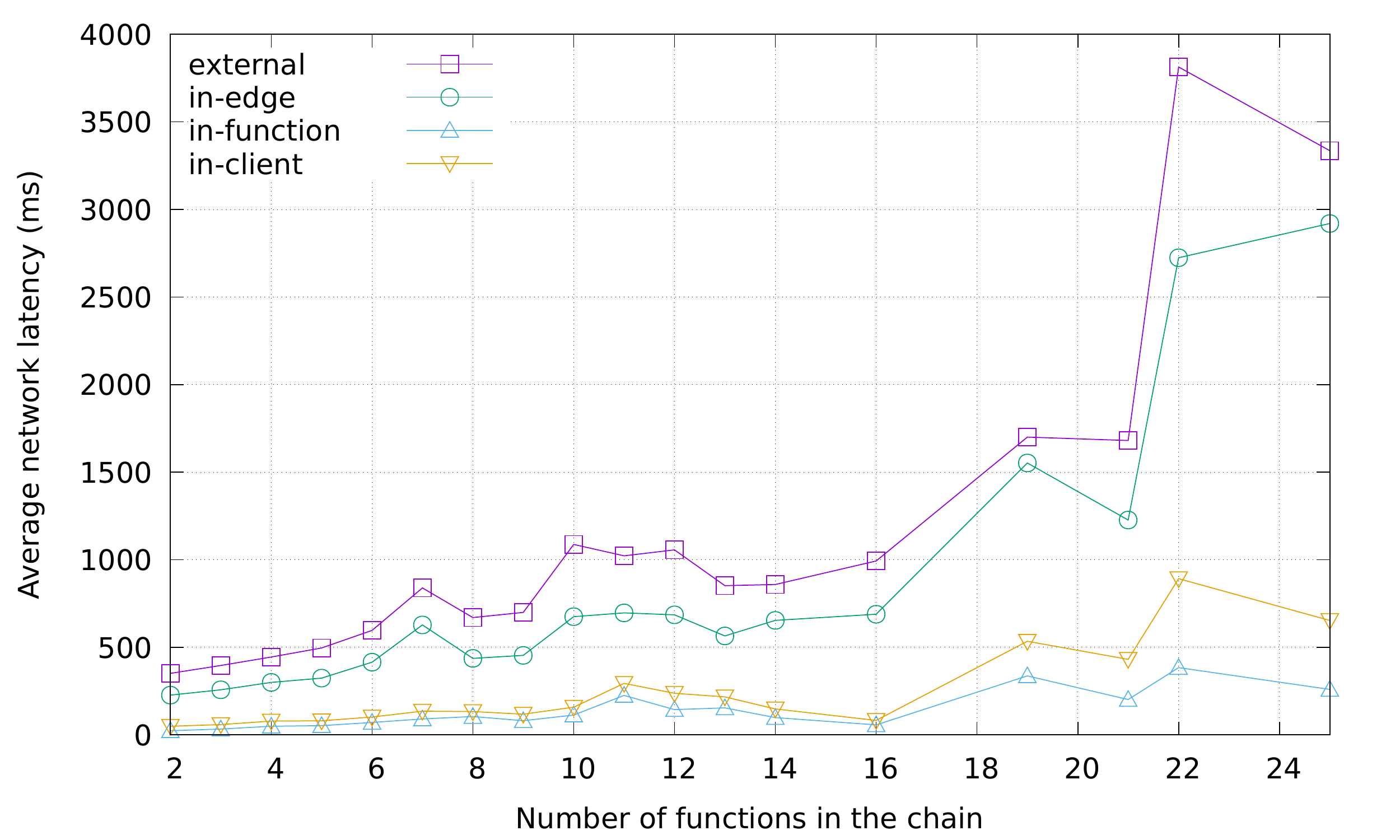}{Network latency, with Load+net
allocation strategy, all execution models, and 500 jobs, per number of
functions in the chain of invocations.}

We conclude with a deeper insight on the latency due to the network
transfers only in \rfig{002-pcs-netlat-iiot-procnet-500}, combining
together jobs with the same chain length.
The sharp increase of all curves for longer function chains is because they
tend to be more heavyweight in the Sp\r{a}r-generated traces, in accordance
with the real ones from Alibaba.
With all the execution models the average latency increases with
the chain length; however, such an increase is more evident with the
\textit{external} and \textit{in-edge} models.
Finally, the \textit{in-client} strategy deviates in a noticeable
manner from \textit{in-function} only for the jobs with longest chains.
  \section{Conclusions and outlook}%
  \label{sec:conclusions}%
  In this paper we have provided an overview of the nascent topic of
stateful \ac{FaaS} to support in-network computing.
%
We have identified four reference execution models: \textit{external},
where persistence is provided by an external service in the cloud
(current standard); \textit{in-edge}, which envisions
distributing the data within the edge nodes themselves;
\textit{in-function}, where the state remains into a stateful
instance of the function's container; and, \textit{in-client}, where
the client's device is the sole owner of the application's state.
The four models have different implications on the architecture and protocols,
and we envisage that further research is required to determine which one
is best for a given deployment or set of edge applications.
As an initial step in this direction, we have built a trace-driven simulator
to identify the traffic overhead and latency.
Our results have shown that \textit{in-client} is very promising
since it can surpass \textit{external} and \textit{in-edge}, while
providing users with almost same performance as \textit{in-function},
despite being simpler and more amenable to system-wide optimization.
%


\begin{thebibliography}{10}

\bibitem{Campbell2019}
Mark Campbell.
\newblock {Smart Edge : The Center of Data Gravity Out of the Cloud}.
\newblock {\em Computer}, 52(December):99--102, 2019.

\bibitem{Ramachandran2021}
Umakishore Ramachandran, Harshit Gupta, Adam Hall, Enrique Saurez, and Zhuangdi
  Xu.
\newblock {A Case for Elevating the Edge to be a Peer of the Cloud}.
\newblock {\em GetMobile: Mobile Computing and Communications}, 24(3):14--19,
  2021.

\bibitem{Liu2020}
Yaqiong Liu, Mugen Peng, Guochu Shou, Yudong Chen, and Siyu Chen.
\newblock {Toward Edge Intelligence: Multiaccess Edge Computing for 5G and
  Internet of Things}.
\newblock {\em IEEE Internet of Things Journal}, 7(8):6722--6747, 2020.

\bibitem{Sapio2017}
Amedeo Sapio, Ibrahim Abdelaziz, Abdulla Aldilaijan, Marco Canini, and Panos
  Kalnis.
\newblock {In-Network Computation is a Dumb Idea Whose Time Has Come}.
\newblock {\em Proc. ACM HotNets 2021}.

\bibitem{Castro:2019:RSC:3372896.3368454}
Paul Castro, Vatche Ishakian, Vinod Muthusamy, and Aleksander Slominski.
\newblock {The Rise of Serverless Computing}.
\newblock {\em Commun. ACM}, 62(12):44--54, nov 2019.

\bibitem{Baldini2017}
Ioana Baldini, Perry Cheng, Stephen~J. Fink, Nick Mitchell, Vinod Muthusamy,
  Rodric Rabbah, Philippe Suter, and Olivier Tardieu.
\newblock {The serverless trilemma: Function composition for serverless
  computing}.
\newblock {\em Proc. Onward! 2017}.

\bibitem{Cicconetti2020}
Claudio Cicconetti, Marco Conti, Andrea Passarella, and Dario Sabella.
\newblock {Toward Distributed Computing Environments with Serverless Solutions
  in Edge Systems}.
\newblock {\em IEEE Communications Magazine}, 58(3):40--46, mar 2020.

\bibitem{Bhasi2021}
Vivek~M Bhasi, Jashwant~Raj Gunasekaran, Prashanth Thinakaran, Cyan~Subhra
  Mishra, Mahmut~Taylan Kandemir, and Chita Das.
\newblock {Kraken: Adaptive Container Provisioning for Deploying Dynamic DAGs
  in Serverless Platforms}.
\newblock {\em Proc. ACM SoCC 2021}.

\bibitem{Rausch2021}
Thomas Rausch, Alexander Rashed, and Schahram Dustdar.
\newblock {Optimized container scheduling for data-intensive serverless edge
  computing}.
\newblock {\em Future Generation Computer Systems}, 114:259--271, 2021.

\bibitem{Mvondo2021}
Djob Mvondo, E~N~S Lyon, Kevin Nguetchouang, Lucien Ngale, E~N~S Lyon, Renaud
  Lachaize, Tim Wood, Daniel Hagimont, and Alain Tchana.
\newblock {OFC : An Opportunistic Caching System for FaaS Platforms}.
\newblock {\em Proc. ACM EuroSys 2021.}

\bibitem{Khandelwal2021}
Anurag Khandelwal, Joao Carreira, Neeraja~J Yadwadkar, Raluca A D~A Popa,
  Joseph~E Gonzalez, I~O~N Stoica, and David~A Patterson.
\newblock {What Serverless Computing Is and Should Become: The Next Phase of
  Cloud Computing}.
\newblock {\em Communications of the ACM}, 64(5):76--84, 2021.

\bibitem{Aslanpour2021}
Mohammad~S. Aslanpour, Adel~N. Toosi, Claudio Cicconetti, Bahman Javadi, Peter
  Sbarski, Davide Taibi, Marcos Assuncao, Sukhpal~Singh Gill, Raj~K Gaire, and
  Schahram Dustdar.
\newblock {Serverless Edge Computing: Vision and Challenges}.
\newblock {\em Proc. AusPDC 2021.}

\bibitem{Ascigil2021}
Onur Ascigil, Argyrios Tasiopoulos, Truong~Khoa Phan, Vasilis Sourlas, Ioannis
  Psaras, and George Pavlou.
\newblock {Resource Provisioning and Allocation in Function-as-a-Service
  Edge-Clouds}.
\newblock {\em IEEE Transactions on Services Computing}, 1374(c):1--14, 2021.

\bibitem{Wang2020b}
Lin Wang, Lei Jiao, Ting He, Jun Li, and Henri Bal.
\newblock {Service Placement for Collaborative Edge Applications}.
\newblock {\em IEEE/ACM Transactions on Networking}, pages 1--14, 2020.

\bibitem{9193994}
Claudio Cicconetti, Marco Conti, and Andrea Passarella.
\newblock {A Decentralized Framework for Serverless Edge Computing in the
  Internet of Things}.
\newblock {\em IEEE Transactions on Network and Service Management}, pages
  1--1, 2020.

\bibitem{Baresi2019}
L.~Baresi, D.~F. Mendon{\c{c}}a, M.~Garriga, S.~Guinea, and G.~Quattrocchi.
\newblock {A unified model for the mobile-edge-cloud continuum}.
\newblock {\em ACM Transactions on Internet Technology}, 19(2), 2019.

\bibitem{Mohan2019}
Anup Mohan, Harshad Sane, Kshitij Doshi, Saikrishna Edupuganti, Naren Nayak,
  and Vadim Sukhomlinov.
\newblock {Agile cold starts for scalable serverless}.
\newblock {\em Proc. USENIX HotCloud 2019.}

\bibitem{Puliafito2020}
Carlo Puliafito, Antonio Virdis, and Enzo Mingozzi.
\newblock {The Impact of Container Migration on Fog Services as Perceived by
  Mobile Things}.
\newblock In {\em Proc. IEEE SMARTCOMP 2020.}

\bibitem{Tian2019}
Huangshi Tian, Yunchuan Zheng, and Wei Wang.
\newblock {Characterizing and Synthesizing Task Dependencies of Data-Parallel
  Jobs in Alibaba Cloud}.
\newblock {\em Proc. ACM SoCC 2019.}

\bibitem{Persico2017}
Valerio Persico, Alessio Botta, Pietro Marchetta, Antonio Montieri, and Antonio
  Pescap{\'{e}}.
\newblock {On the performance of the wide-area networks interconnecting
  public-cloud datacenters around the globe}.
\newblock {\em Computer Networks}, 112:67--83, jan 2017.

\end{thebibliography}



\begin{acronym}
  \acro{3GPP}{Third Generation Partnership Project}
  \acro{5G-PPP}{5G Public Private Partnership}
  \acro{AA}{Authentication and Authorization}
  \acro{AI}{Artificial Intelligence}
  \acro{API}{Application Programming Interface}
  \acro{AP}{Access Point}
  \acro{AR}{Augmented Reality}
  \acro{BGP}{Border Gateway Protocol}
  \acro{BSP}{Bulk Synchronous Parallel}
  \acro{BS}{Base Station}
  \acro{CDF}{Cumulative Distribution Function}
  \acro{CFS}{Customer Facing Service}
  \acro{CPU}{Central Processing Unit}
  \acro{DAG}{Directed Acyclic Graph}
  \acro{DHT}{Distributed Hash Table}
  \acro{DNS}{Domain Name System}
  \acro{ETSI}{European Telecommunications Standards Institute}
  \acro{FCFS}{First Come First Serve}
  \acro{FSM}{Finite State Machine}
  \acro{FaaS}{Function as a Service}
  \acro{GPU}{Graphics Processing Unit}
  \acro{HTML}{HyperText Markup Language}
  \acro{HTTP}{Hyper-Text Transfer Protocol}
  \acro{ICN}{Information-Centric Networking}
  \acro{IETF}{Internet Engineering Task Force}
  \acro{IIoT}{Industrial Internet of Things}
  \acro{IPP}{Interrupted Poisson Process}
  \acro{IP}{Internet Protocol}
  \acro{ISG}{Industry Specification Group}
  \acro{ITS}{Intelligent Transportation System}
  \acro{ITU}{International Telecommunication Union}
  \acro{IT}{Information Technology}
  \acro{IaaS}{Infrastructure as a Service}
  \acro{IoT}{Internet of Things}
  \acro{JSON}{JavaScript Object Notation}
  \acro{LCM}{Life Cycle Management}
  \acro{LL}{Link Layer}
  \acro{LTE}{Long Term Evolution}
  \acro{MAC}{Medium Access Layer}
  \acro{MBWA}{Mobile Broadband Wireless Access}
  \acro{MCC}{Mobile Cloud Computing}
  \acro{MEC}{Multi-access Edge Computing}
  \acro{MEH}{Mobile Edge Host}
  \acro{MEPM}{Mobile Edge Platform Manager}
  \acro{MEP}{Mobile Edge Platform}
  \acro{ME}{Mobile Edge}
  \acro{ML}{Machine Learning}
  \acro{MNO}{Mobile Network Operator}
  \acro{NAT}{Network Address Translation}
  \acro{NFV}{Network Function Virtualization}
  \acro{NFaaS}{Named Function as a Service}
  \acro{OSPF}{Open Shortest Path First}
  \acro{OSS}{Operations Support System}
  \acro{OS}{Operating System}
  \acro{OWC}{OpenWhisk Controller}
  \acro{PMF}{Probability Mass Function}
  \acro{PU}{Processing Unit}
  \acro{PaaS}{Platform as a Service}
  \acro{PoA}{Point of Attachment}
  \acro{QoE}{Quality of Experience}
  \acro{QoS}{Quality of Service}
  \acro{RPC}{Remote Procedure Call}
  \acro{RR}{Round Robin}
  \acro{RSU}{Road Side Unit}
  \acro{SBC}{Single-Board Computer}
  \acro{SDN}{Software Defined Networking}
  \acro{SJF}{Shortest Job First}
  \acro{SLA}{Service Level Agreement}
  \acro{SMP}{Symmetric Multiprocessing}
  \acro{SoC}{System on Chip}
  \acro{SRPT}{Shortest Remaining Processing Time}
  \acro{SPT}{Shortest Processing Time}
  \acro{STL}{Standard Template Library}
  \acro{SaaS}{Software as a Service}
  \acro{TCP}{Transmission Control Protocol}
  \acro{TSN}{Time-Sensitive Networking}
  \acro{UDP}{User Datagram Protocol}
  \acro{UE}{User Equipment}
  \acro{URI}{Uniform Resource Identifier}
  \acro{URL}{Uniform Resource Locator}
  \acro{UT}{User Terminal}
  \acro{VANET}{Vehicular Ad-hoc Network}
  \acro{VIM}{Virtual Infrastructure Manager}
  \acro{VR}{Virtual Reality}
  \acro{VM}{Virtual Machine}
  \acro{VNF}{Virtual Network Function}
  \acro{WLAN}{Wireless Local Area Network}
  \acro{WMN}{Wireless Mesh Network}
  \acro{WRR}{Weighted Round Robin}
  \acro{YAML}{YAML Ain't Markup Language}
\end{acronym}

\end{document}